\documentclass[conference]{IEEEtran}
\IEEEoverridecommandlockouts
\usepackage{cite}
\usepackage{algorithmic}
\usepackage{graphicx}
\usepackage{textcomp}
\usepackage{xcolor}
\usepackage{caption}
\usepackage{multirow}
\usepackage{subcaption}
\usepackage{listings}
\usepackage{float}
\usepackage{lipsum}
\usepackage[numbers]{natbib}
\usepackage{xcolor}
\usepackage[T1]{fontenc}

\usepackage{tikz}
\usepackage{comment}
\usepackage{color}
\usepackage{xspace}
\usepackage{array}
\usepackage{arydshln}
\usepackage{amsmath,amssymb,amsfonts,bbm}
\usepackage{xcolor}

\makeatletter

\def\BibTeX{{\rm B\kern-.05em{\sc i\kern-.025em b}\kern-.08em
    T\kern-.1667em\lower.7ex\hbox{E}\kern-.125emX}}
    
\DeclareRobustCommand\onedot{\futurelet\@let@token\@onedot}
\def\@onedot{\ifx\@let@token.\else.\null\fi\xspace}

\def\eg{\emph{e.g}\onedot} 
\def\ie{\emph{i.e}\onedot} 
 
\def\etc{\emph{etc}\onedot} 
 
\def\etal{\emph{et al}\onedot} 
\makeatother

\begin{document}
\title{API-Miner: an API-to-API Specification Recommendation Engine}
\titlerunning{API-Miner: an API-to-API Specification Recommendation Engine}
\author{
Sae Young Moon,
Gregor Kerr,
Fran Silavong,
Sean Moran
}

\institute{JP Morgan Chase\\ \email{\footnotesize{
\{saeyoung.moon, gregor.kerr, fran.silavong, sean.j.moran\}@jpmchase.com
}}}

\maketitle{}      
\begin{abstract}
When designing a new API for a large project, developers need to make smart design choices so that their code base can grow sustainably. To ensure that new API components are well designed, developers can learn from existing API components. However, the lack of standardized methods for comparing API designs makes this learning process time-consuming and difficult. To address this gap we developed API-Miner, to the best of our knowledge, one of the first API-to-API specification recommendation engines. API-Miner retrieves relevant specification components written in OpenAPI (a widely adopted language used to describe web APIs). API-miner presents several significant contributions, including: (1) novel methods of processing and extracting key information from OpenAPI specifications, (2) innovative feature extraction techniques that are optimized for the highly technical API specification domain, and (3) a novel log-linear probabilistic model that combines multiple signals to retrieve relevant and high quality OpenAPI specification components given a query specification. We evaluate API-Miner in both quantitative and qualitative tasks and achieve an overall of 91.7\% recall@1 and 56.2\% F1, which surpasses baseline performance by 15.4\% in recall@1 and 3.2\% in F1. Overall, API-Miner will allow developers to retrieve relevant OpenAPI specification components from a public or internal database in the early stages of the API development cycle, so that they can learn from existing established examples and potentially identify redundancies in their work. It provides the guidance developers need to accelerate development process and contribute thoughtfully designed APIs that promote code maintainability and quality. Code is available on GitHub at \url{https://github.com/jpmorganchase/api-miner}.
\end{abstract}

\begin{IEEEkeywords}
API contract, API specification, structured document matching, recommendation system
\end{IEEEkeywords}

\section{Introduction}

The Web continues to grow at an incredibly fast pace, with more than an average of 2,000 new web APIs (Application Programming Interface) being added per year since 2015~\cite{apistat}. A Web API is an interface that provides access to the functionalities of a software service, and these functionalities can be accessed via the endpoints (\ie~URLs) \cite{apievolution}. Given the task to develop a new API or extend the functionalities of an existing one, developers often adopt the `API-first' approach~\footnote{\url{https://swagger.io/resources/articles/adopting-an-api-first-approach/}}. In this approach, developers begin by writing a specification to document the APIs design, such as the capabilities of each endpoint. This ensures that the API design is sound and consistent with the rest of the project before developers invest time to write code. 

\begin{figure*}[htp]
    \centering
    \includegraphics[width=17cm]{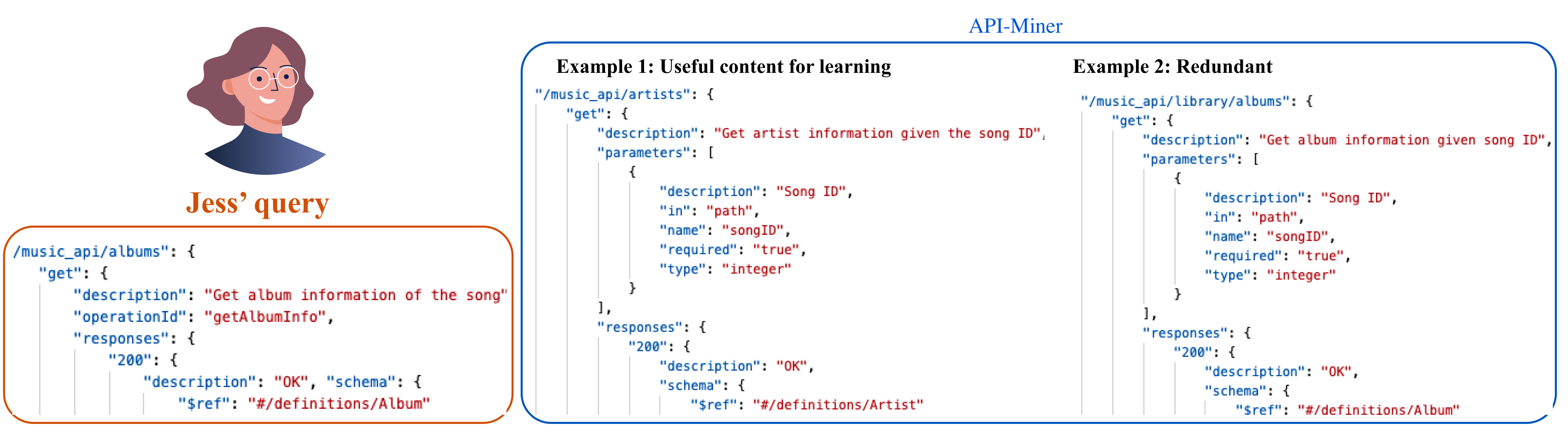}
    \caption{Visualization of the motivational example, showing how API-Miner's results can help developers with API design}
    \label{fig:api_spector_example}
\end{figure*}

Designing new API functionality can be a daunting task. To visualize the current challenges, we will go through a motivational example inspired by conversations that we had with web API developers at our company. Suppose that Jess is a software developer who is working on a Music Library API and is currently adding new functionality that retrieves the album information of a song. Following the `API-first' approach, she starts by adding a new endpoint to the APIs specification. Jess can write some of the sections with ease, however still encounters some roadblocks. What input parameters should she consider? Is there a better way to implement this?

To help developers like Jess make smarter design choices, we developed API-Miner, an API-to-API Specification Recommendation Engine that takes the user's endpoint specification draft (\ie~component of a specification that corresponds to a specific endpoint) as the input, and retrieves relevant and high quality examples from the database. In Jess' example shown in Figure~\ref{fig:api_spector_example}, API-Miner can return examples that are useful for learning, such as an endpoint that gets artist information given song ID. Or it could return a redundant endpoint, letting her know that this functionality has already been developed. In short, API-Miner supports developers in every step of their API design journey. As they refine their specification draft, API-Miner will be able to recommend more relevant examples to solidify their API design. For the purposes of this paper, we focused on recommending OpenAPI specification components~\cite{swagger} because they are a popular standard for documenting RESTful APIs~\cite{restfulapi}. 

The main contributions of this paper are as follows:

\begin{itemize}
\item \textbf{API-to-API Specification Recommendation Engine}: We design a recommendation engine that retrieves relevant endpoint specifications given the user's endpoint specification query. `Endpoint specification' refers to the section of an OpenAPI specification that corresponds to an endpoint of interest, and that relevant endpoint specifications are defined as those that: (1) provide additional content not present in the users current specification and/or (2) are redundant in structure and content with respect to the user's query. Furthermore, API-Miner can recommend similar endpoints for syntactically invalid endpoint specifications (\ie~missing required properties or non-finished textual descriptions).

\item \textbf{Featurization}: We propose to represent endpoint specifications as a \textit{tree} to fully utilize the hierarchical relationship between various sections and extract features from individual nodes within the tree and its path. This addresses situations where the semantic meaning of a token differs depending on its location within the hierarchy and/or its relation to its parent node. Our method extracts three set of features to better characterize endpoint specifications: (1) Tree Path tokens capturing the hierarchical relationship, (2) Natural Language Tokens containing domain-specific terms (keyword tokens) and (3) the API endpoint name itself.

\item \textbf{Multi-Signal Ranking Model}: We developed a probabilistic log-linear model to fuse multiple signals of similarity obtained from our proposed featurization methods, along with the endpoint's quality metrics, to retrieve a ranked list of relevant and high quality endpoint specifications. We show that the retrieval performance improves when more than one source of the similarity is considered.
\end{itemize}

The remainder of this paper is organized as follows: Section~\ref{sec:related_work}, we describe the related work in the API-to-API recommendation literature. In Section~\ref{sec:method}, we describe our contribution, an API Specification Recommendation Engine called \emph{API-Miner}. In Section~\ref{sec:eval}, we experimentally evaluate API-Miner on 1,000 random queries generated from the corpus and a pooling-based user study. Finally, we draw conclusions and provide pointers for future work in Section~\ref{sec:conclusions}.

\section{Related Work}
\label{sec:related_work}

\subsection{API Retrieval Systems} 
API-Miner helps developers by retrieving endpoint specifications that are relevant to the endpoint they are currently designing. To the best of our knowledge, there is no existing work that retrieves relevant specification components given an endpoint of interest. Therefore, our baselines had to be derived from related works that retrieve entire APIs given a user query as outlined in this section. Broadly, the APIs can be represented as: (1) keywords or text, or (2) a specification. We will review related work that processes both API representations in this section. 

\textbf{Representing APIs with keywords or text}: Web API recommendation systems developed for mash development typically represent APIs by their keywords or by text descriptions. They analyze these API representations, often using compatibility graphs, to retrieve a set of compatible APIs that developers can `mash' together to make a software product. Notable works include: Qi~\etal's weighted API correlation graph search algorithm \cite{qifindingallyouneed}, Fletcher's regularized user preference embedded matrix factorization recommendation algorithm \cite{fletcherregularizingmatrix}, Wang~\etal's random walk on knowledge graph algorithm \cite{wangrandomwalk}, and Thung~\etal's TF-IDF-based vector semantic similarity algorithm \cite{thungwebapirec}. Ultimately, the above approaches were not considered as baselines for API-Miner because their use case is significantly different from ours. 

\textbf{Representing APIs with a specification}: There exists web API retrieval systems that represent APIs by their specification, such as Web Application Description Language (WADL) \cite{hadley2006web}, Web Services Description Language (WSDL) \cite{wsdl}, Universal Description, Discovery and Integration (UDDI) \cite{uddi}, and OpenAPI Specification \cite{openapispec}. Since API-Miner processes OpenAPI specification components, we derived our baselines based on these retrieval methods.

\begin{figure*}[htp]
    \centering
    \includegraphics[width=17cm]{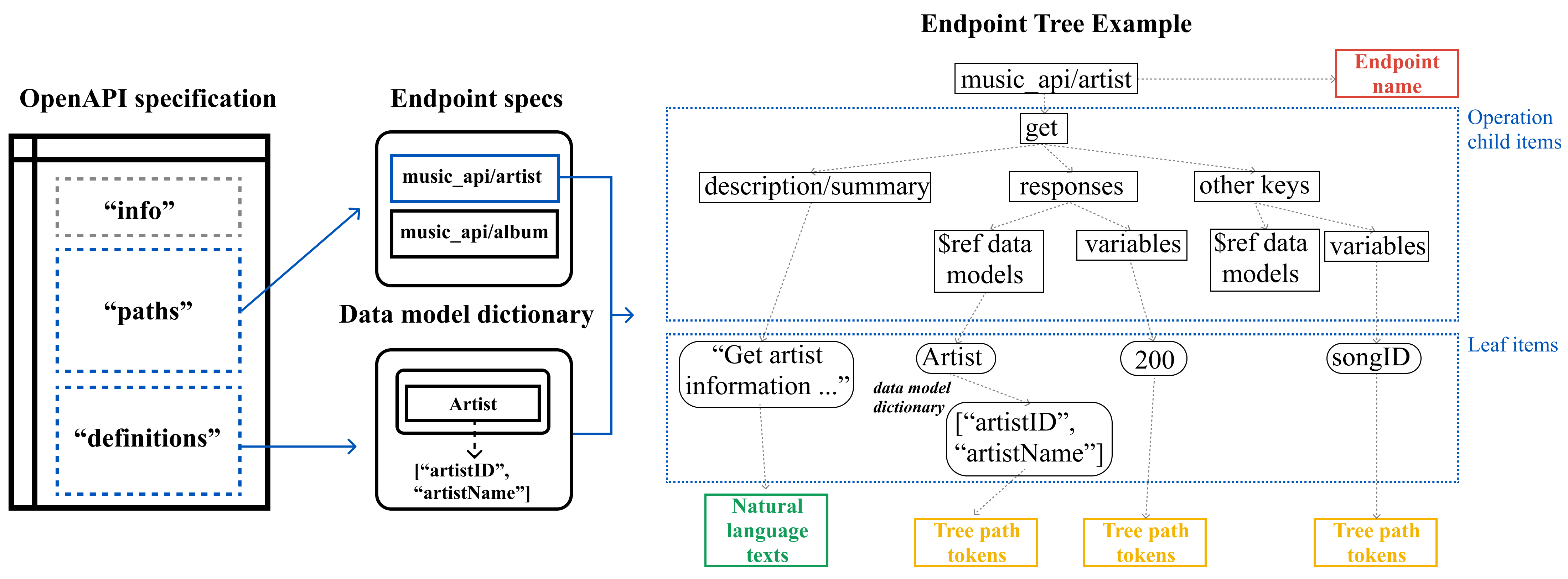}
    \caption{Tree visualization of a endpoint specification}
    \label{fig:spec_tree}
\end{figure*}

String matching has been explored to evaluate the similarity of API specifications written in WADL and WSDL (\cite{khorasgani2011web}, \cite{bai2011fuzzy}, \cite{wsdlstructuralsim}, \cite{uddisimilarity}), for the purpose of retrieving relevant APIs. Within the domain of OpenAPI specification, Peng~\etal. \cite{openapifuzzy} proposed to leverage the Levenshtein distance of the descriptive (\eg~text) and functional (\eg~data-type) properties found in the service and resource levels of a specification. The similarity of specifications was determined either by a weighted combination of the distances or by assigning a relevance label based on the distance (\eg~low/middle/high). However, string matching approaches are limited because they cannot capture semantic meaning, rendering them unable to retrieve semantically similar specifications that do not contain exact string matches. 


To address the limitations of string matching algorithms, various works leveraged textual description present in specifications to compute semantic similarity and retrieve relevant APIs. For instance, Wang \etal. \cite{wsdlsimilarity} developed a specification-to-specification search engine by computing the semantic and structural similarity of specifications written in WSDL. Semantic similarity was computed based on WordNet embeddings of the natural language descriptions, and structural similarity was computed based on data types and parameters present in the specification. Overall, their semantic similarity solution reported 41\% recall@1 and their structural similarity solution reported 22\% recall@1, where recall@1 is the percentage of times the system was able to retrieve the best match as its first recommendation. Although Wang \etal's implementation focuses on entire API specifications, this could not be used as a direct baseline without significant re-invention due to our approach focusing on endpoint components specifically. However, we adopted their best performing API retrieval approach to develop a semantic similarity baseline for our study. To the best of our knowledge, there has been limited application of this technique in the domain of OpenAPI specification.

\subsection{Vector Semantics in Low Resource Domains} 
\label{sec:vector_semantics_low_resource}

Domain specific terminology used in API specifications poses two main challenges for semantic representation learning: (1) technical terms are growing and non-standarized by nature (\eg~acronyms), and (2) standard pre-trained language model embeddings are unable to capture the semantic meaning behind technical terms. Although works have been explored to fine-tune embeddings for low-resource domains, they are difficult to achieve with an ever growing vocabulary as described in challenge (1) \cite{rareembeddingsonfly}\cite{rarewordsattentionmimick}\cite{mimick}\cite{wces}. Positive Pointwise Mutual Information (PPMI) is used to compare the probability of two events occurring together and the probability of these two events occurring independently. Previous work \cite{ppmidirichlet} proposed to leverage PPMI to represent low resource words, however, it is limited due to PPMI not leveraging rich contextual representations that can be obtained with pre-trained language models. Therefore in this work, we explore methods of obtaining enriched representations of text that leverage contextual understanding of pre-trained language models, while also sufficiently representing technical terms. 

\section{Proposed Methods}\label{sec:method}

\begin{figure*}[htp]
    \centering
    \includegraphics[width=16cm]{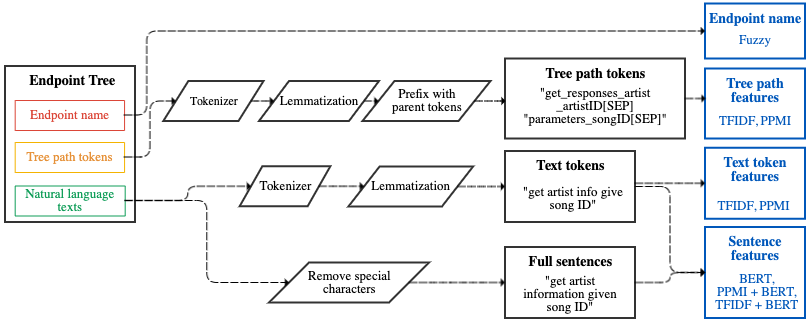}
    \caption{Featurization of endpoint specifications}
    \label{fig:featurization}
\end{figure*}

\subsection{OpenAPI Specification Vector Representation}
API-Miner expects endpoint specifications (\ie~components of an OpenAPI specification that correspond to an endpoint of interest) as the user query. The ability to process specification queries, as opposed to simple keywords or text descriptions, brings several advantages. Firstly, a specification can represent complex design choices, such as input parameters and responses, that are difficult and time-consuming to translate into textual descriptions. Secondly, a developer naturally builds upon their design by iterating on their specification, often starting with only the endpoints. The ability to recursively query using their current specification draft would optimize their productivity and efficiency. 


OpenAPI specifications can be written in JSON or YAML, containing sections for describing web APIs in a standardized and comprehensive manner \cite{swagger}. On a high level, a specification can be split into 3 major sections: (1) the `info' section contains general information about the API, (2) the `paths' section contains endpoint specifications that describe functionalities of specific endpoints, and (3) the `definitions' section outlines the common data models that are re-used throughout the API. The focus of our work is to compute the similarity of specifications on an endpoint level, and thus only the `paths' and `definitions' sections are considered. Figure \ref{fig:spec_tree}, illustrates that the `paths' section is transformed to form endpoint specifications, and the `definitions' section is parsed to create a dictionary, where the keys represent the data model names and the values represent their properties (\ie~input and outputs of the data model).

We propose to combine information from the `paths' and `definitions' sections to represent each endpoint in tree form, as shown in Figure \ref{fig:spec_tree} to truly capture the rich information present in the hierarchical structure of a specification and offer context-aware recommendation. For instance, a word used in a parameter does not have the same representation as one in the data model, which enables more targeted recommendation given a user query.  The top level item of the tree represents the endpoint name, and its child items represent the operations defined under the endpoint. The leaf items contain key content such as the description text, variable names and properties of data models that were referenced using the \emph{$\$ref$} token. During pre-processing, two types of key components are extracted from the leaf items of an endpoint tree:

\begin{enumerate}
  \item Variable names (\eg parameters, response names, properties) that are used to generate tree path tokens. These variable names are prefixed with relevant parent tokens to form tree path tokens (\eg~`songID' is represented as `parameters\_songID').
  \item Natural language text (\ie summaries and descriptions of the operations). Note that descriptions of the responses or parameters were not considered, because they are often standardized (\eg~`OK' for response 200) and therefore would not be helpful for characterizing the endpoint. 
\end{enumerate}

To prepare the database for API-Miner, a collection of endpoint trees are obtained from the OpenAPI specifications available. Endpoint names, tree path tokens and natural language texts contained in each endpoint tree are processed to generate features that characterize every unique endpoint. This allows API-Miner to retrieve endpoint specifications that are relevant to the query (\ie~user's current endpoint specification draft). This process is detailed in Section \ref{sec:probmodel}.

\subsection{Probabilistic Log-Linear Model for API Endpoint Ranking}\label{sec:probmodel}

Although guidelines for best practises exist, there is often a high variability in the way OpenAPI specifications are written due to differences in writing style. For example, some developers mostly focus on descriptions, while others simply outline inputs and outputs. In order to account for this variability, we developed a probabilistic log-linear model that fuses multiple similarity signals. Probabilistic log linear models have found success in domains such as Natural Language Processing (NLP)~\cite{Yang13,och-ney-2002-discriminative,Papineni1997FeaturebasedLU} and Information Retrieval (IR)~\cite{Gysel16}, particularly point-wise Learning To Rank (LTR)~\cite{Liu09}. Log-linear models encourage a rich variety of feature representations to influence the probabilistic estimation of relevance, making it ideal for our endpoint ranking task. We define the probability $P(e_{i}|e_{j})$ of a match between two endpoint specifications in Equation~\ref{eq:fusion1}, which is a linear combination of features and weights coming from \emph{tree path}, \emph{fuzzy string}, description \emph{text} and endpoint specification \emph{quality}.
\vspace{-0.1in}

\begin{equation}
    \label{eq:fusion1}
    P(e_{i}|e_{j};w) = \frac{\exp (\sum^{F-1}_{k=1} w_{k}f_{k}(e_{i}, {e_{j})} + w_{F}f_{q}(e_{j}))}{\sum^{N_{E}}_{i=1} \exp( \sum^{F-1}_{k=1} w_{k}f_{k}(e_{i},e_{j})+ w_{F}f_{q}(e_{i}))}
\end{equation}



\noindent{$e_{i}, e_{j}$ is the collection of operations (Figure~\ref{fig:spec_tree}) associated with endpoints $i$ and $j$. The $f_{k}(.)$ are $F$ \emph{feature functions} that measure different aspects of the endpoints based on properties of their operations (text, tree path token, endpoint name), and $f_{q}(.)$ is the quality bias. $\left\{w_{1},w_{2},w_{3}, w_{4}\right\}$ denote the fusion weights.}

\begin{equation}
    \label{eq:fusion2}
    \hat{P}(e_{i}|e_{j}) = \frac{P(e_{i}|e_{j})}{\max_{k} P(e_{k}|e_{j})}
\end{equation}

\noindent{For inference, we apply Equation~\ref{eq:fusion2} to rank candidate endpoints $e_{i}$ given a query endpoint $e_{j}$. This ranking is achieved by computing $\hat{P}(e_{i}|e_{j})$ for all endpoints $e_{i}$ in the database and sorting the resulting probabilities in descending order. To ensure better readability of small probabilities without changing the ranked order, we normalize the probabilities by the maximum observed matching probability.}

\begin{figure*}[htp]
    \centering
    \includegraphics[width=15cm]{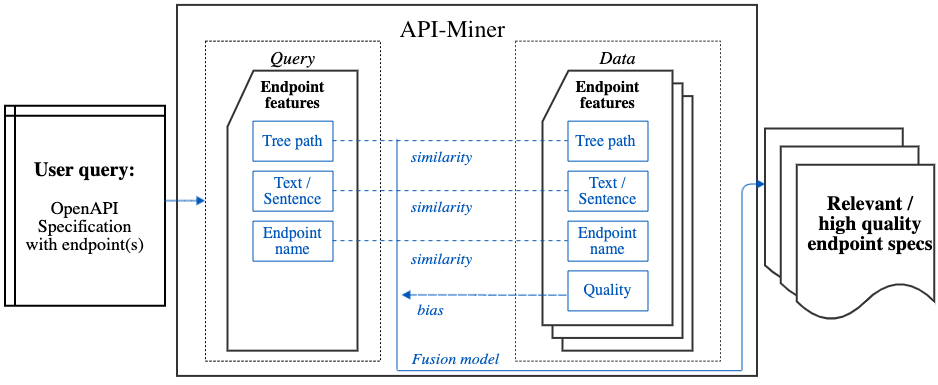}
    \caption{Overall diagram of API-Miner}
    \label{fig:overall_diagram}
\end{figure*}

In our instantiation of the API-Miner log-linear probabilistic model, we define four ($F{=}4$) feature functions. Three feature functions measure similarity between query and data endpoint specifications and these form our \emph{query-dependent} features. These similarities are measured between: 1) tree path tokens (Section~\ref{sec:treepath}), 2) endpoint names (Section~\ref{sec:natlang}) and 3) texts describing operations (Section~\ref{sec:fuzzy}). We also add a fourth feature function that is \emph{query-independent}, which rates the quality of specifications in the database (Section~\ref{sec:quality}). In principle, the fusion weights $\left\{w_{1},w_{2},w_{3}, w_{4}\right\}$ could be learned by minimising the negative log-likelihood via gradient descent, $\hat{w}=argmin_{w}-\sum^{N}_{i=1}\log P(e_{i}|e^{'}_{i};w)$ based on a supervised signal of $N$ relevant endpoint pairs $(e_{i},e^{'}_{i})$. However, due to the lack of large-scale annotated data, we heuristically set the weights of the quality features ($f_q$) to be 0.1 and then distribute the remaining 0.9 of weights evenly among the rest of the features to emphasize relevance of retrieved endpoint. In Sections \ref{sec:treepath} to \ref{sec:quality}, we summarize the different approaches that we explored to evaluate the similarity and quality of endpoint specifications. The feature extraction methods, as well as the overall API-Miner work flow is outlined in Figures \ref{fig:featurization} and \ref{fig:overall_diagram}.

\subsubsection{Tree path token similarity}\label{sec:treepath}

We consider similarity of tree path tokens for our probabilistic model. Tree path tokens - generated from variables tokens, were extracted from endpoint trees by preprocessing (\ie~tokenizing, removing special characters, lemmatizing) parameters and properties of referenced data models. A variable token is prefixed with its parent node name or a combination of the operation and response names to obtain a tree path token such as \emph{`parameters\_songID'} or \emph{`get\_responses\_200\_Artist\_artistName'}. We denote $e^{tree}$ as the collection of tree path tokens $t$ in endpoint $e$. We experiment with two different techniques, firstly, Term Frequency - Inverse Document Frequency TF-IDF~\cite{Sparck88} which is used to count frequency of words to determine how relevant those words are to a given document and secondly PMI~\cite{Church90}, where for both experiments we convert the collection of tree path tokens to a numeric feature vector for endpoint similarity comparison. The $TF{-}IDF$ weight for token $t_{i}$ in endpoint $e_{j}$ is defined in Equation~\ref{eq:tfidf}.

\begin{equation}
    \label{eq:tfidf}
    \begin{split}
        \text{TF{-}IDF}(t_{i}, e^{tree}_{j}) &= (\frac{\#(t_{i}, e^{tree}_{j})} {\sum_{t_{k}\in V_{T}} \#(t_{k}, e^{tree}_{j})}\\
        &=\times  \log(\frac{N_{E}}{\sum_{j\in V_{E}} \mathbbm{1}_{(\#(t_{i}, e^{tree}_{j})\ge 1)}})
    \end{split}
\end{equation}

\noindent{where $\#(t_{i},e^{tree}_{j})$ is the count of tree path token $t_{i}$ in endpoint $e_{j}$ and $N_{E}$ is the number of endpoints. We represent an endpoint's tree path tokens $e^{tree}_{j}$ as a vector $x_{j}\in\Re^{N_{T}}$ weighted by TF-IDF, where $x_{j}[i]=TF{-}IDF(t_{i}, e^{tree}_{j})$. To determine the similarity between endpoints' tree path tokens, we compute the cosine similarity between their TF{-}IDF vectors (Equation~\ref{eq:cosine1}). }

\begin{equation}
    \label{eq:cosine1}
   f_{1}(e_{i},e_{j}) = cos(e^{tree}_{i},e^{tree}_{j}) = 
   \frac{x_{i}.x_{j}}{\sqrt{x_{i}^{T}x_{i}}\sqrt{x_{j}^{T}x_{j}}}
\end{equation}

TF{-}IDF feature scoring seeks to promote tokens that appear frequently within an endpoint specification and less frequently across the collection of $N_{E}$ endpoint specifications. A potential issue with TF{-}IDF is that it does not capitalise on token correlation. For example, token \emph{`parameters\_firstName'} should have a higher matching score to a token \emph{`parameters\_lastName'}, as both tokens typically occur in similar context. To extract token correlation, we compute the PPMI score between the $N_{T}$ tree tokens in $V_{T}$. Given endpoints $e\in V_{E}$ and their tree path tokens $t\in V_{T}$ we define the collection of all possible token pairs by $D$. We denote as $\#(t_{i},t_{j})$ as the count of the $j^{th}$ token co-occurring with the $i^{th}$ token in $D$. We further define $\#(t_{i})=\sum_{t_{j}\in V_{T}}\#(t_{i},t_{j})$ and $\#(t_{j})=\sum_{t_{i}\in V_{T}}\#(t_{i},t_{j})$. PPMI can then be expressed as in Equations~\ref{eq:ppmi1}-\ref{eq:ppmi2}.

\begin{equation}
    \label{eq:ppmi1}
    \text{PMI}(t_{i},t_{j}) = \log\frac{P(t_{i},t_{j})}{P(t_{i})P(t_{j})} = \frac{\#(t_{i},t_{j}).|D|}{\#(t_{i})\#(t_{j})}
\end{equation}

\begin{equation}
    \label{eq:ppmi2}
    \text{PPMI}(t_{i},t_{j}) = \max(\text{PMI}(t_{i},t_{j}),0)
\end{equation}

We construct the PPMI matrix $Q\in\Re^{N_{t}\times N_{t}}$, where $Q_{ij}=\text{PPMI}(t_{i},t_{j})$. We leverage matrix $Q$ in a modified version of the cosine similarity (Equation~\ref{eq:cosine2}) that assigns higher similarity to endpoints that share semantically related tokens as defined by PPMI.

\begin{equation}
    \label{eq:cosine2}
   f_{1}(e_{i},e_{j}) = cos(e^{tree}_{i},e^{tree}_{j})_{Q} = \frac{x_{i}Qx_{j}}{\sqrt{x_{i}Qx_{i}}\sqrt{x_{j}Qx_{j}}}
\end{equation}

\noindent{where $x_{i}, x_{j}$ are vectors of integer tree token frequencies.}

Prior to generating TF{-}IDF and PPMI features, we filter out tree path tokens that appear in fewer than 10 endpoint specifications in the database. This is to remove very unique tokens and prevent overfitting. This step is especially important for PPMI features, as it is known to be biased towards those co-occurrences involving rare terms~\cite{Turney10}. For clarity, throughout this paper, we refer to those models that use features from tree path tokens as \emph{`Models using tree path features'}.

\subsubsection{Natural language text similarity}\label{sec:natlang}

The endpoint can also be characterized by the natural language text found in the \emph{`description'} and \emph{`summary'} sections of the operations. The texts can be represented in three ways: (1) by the entire text, (2) by only the keyword tokens present in the text, or (3) a combination of both. 

Firstly, we can leverage pre-trained language models such as BERT~\cite{bert} and SENT-BERT~\cite{sentbert} to generate representations of the entire text.
Specifically, we can obtain sentence-level embeddings based on the first 512 tokens (due to context length limit of BERT and SENT-BERT). This approach allows contextual representations of the text to be captured. SENT-BERT consists of a siamese architecture containing 2 BERT architectures that share the same weights and are closely identical where the difference is that SBERT processes 2 sentences as pairs during training. However due to challenges described in Section \ref{sec:vector_semantics_low_resource}, pre-trained language models struggle with representing low-resource domain-specific terms. 
Secondly, we can obtain keyword tokens by pre-processing the entire text (\ie by tokenizing, removal of stop words and special characters before lemmatizing). We can generate TF{-}IDF and PPMI for these keyword tokens, similar to processes described in Section~\ref{sec:treepath}, which allows us to represent domain specific terms that are not well-represented in pre-trained language models. To prevent overfitting we filter out extremely rare tokens that appear in fewer than 15 endpoints in the database.  

Thirdly, we experiment with combining the above two approaches to obtain `enriched text features'. Simply put, we concatenate the contextual text representations obtained from pre-trained language models with the keyword token representations obtained using  TF{-}IDF or PPMI. Several approaches were explored to linearly project the keyword and entire text features into the same space before concatenation. We experimented with applying truncated SVD to the sparse keyword embeddings, such that the embedding with reduced dimensionality has a total explained variance of 0.95. Then, we applied canonical correlation analysis (CCA) to linearly project the keyword and entire text features into the same space before concatenation. Successful applications of CCA  include fusing multi-modal data~\cite{sargin2007audiovisual} and cross-language tasks in the field of natural language processing \cite{haghighi2008learning}, \cite{vinokourov2002inferring}. The output dimension for the CCA transformation is set to the original dimension of BERT\_SENT (384) or BERT (768). The projection is done by maximizing the correlation (Equation~\ref{eq:cca}) between pre-trained embeddings $X$ and keyword embeddings $Y$. The final features is defined in Equation \ref{eq:ccatransformed}. 

\begin{equation}
    \label{eq:cca}
    (u', v') = arg\max_{u, v} \frac{u^TX^TYv}{\sqrt{(u^TX^TXu)(v^TY^TYv)}}
\end{equation}

\begin{equation}
    \label{eq:ccatransformed}
    e^{text} = concatenate(u'X_, v'Y)
\end{equation}

\noindent{where $u'$ is the projection matrix for pre-trained embeddings $X$, and $v'$ is the projection matrix for keyword embeddings $Y$.} 

We hypothesized that by concatenating the features, we will be able to retain contextual understanding (\ie~with BERT / SENT-BERT features of entire texts), while also representing domain-specific terms sufficiently (\ie~with PPMI / TFIDF features of keyword tokens). These features are also compared using cosine similarity (Equation~\ref{eq:cosine1}) and we represent this similarity as $f_{2}(e_{i},e_{j})=cos(e^{text}_{i},e^{text}_{j})$. Throughout this paper, models that use features from text will be referred to as \emph{`Models using text features'}.

\subsubsection{Fuzzy matching of endpoint names}\label{sec:fuzzy}
The Levenshtein or string edit distance~\cite{LevenshteinSPD66} counts the number of edits (insertions, deletions or substitutions) required on the characters of a string to convert one string to another. The more operations that are required, the greater the distance between the strings. We apply the Levenshtein distance to compute the soft matching between API endpoint names\footnote{The fuzzywuzzy library \cite{fuzzywuzzy} was used to compute the fuzzy match scores.}, where $f_{3}(e_{i},e_{j}){=}lev(e^{name}_{i}, e^{name}_{j})$, and $lev(.)$ is a function that computes the Levenshtein distance. Throughout this paper, models that use fuzzy matching of endpoint names will be referred to as \emph{`Models using fuzzy matching'}.



\subsubsection{Quality score biasing}\label{sec:quality}

\begin{table}[H]
\begin{tabular}{p{1.5cm}|p{1.5cm}|p{4.5cm}}
\hline
\hline
\textbf{Parent key} & \textbf{Required child keys} & \textbf{Expected child key - value data type pairs} \\ \hline

"info" & ["title", "version"] & [("title": str), ("description": str), ("termsOfService": str), ("contact": dict), ("license": dict), ("version": str)]\\ \hline

"get", "put", "post", "delete", "options", "head", "patch" & ["responses"] & [("tags": list), ("summary": str), ("description": str), ("externalDocs": dict), ("operationId": str), ("consumes": list), ( "produces": list), ("parameters": dict), ("responses": dict), ("schemes": list), ("deprecated": bool), ("security": dict)]\\ 

\hline
\hline

\end{tabular}
\caption{\label{tab:quality_table} Table to derive quality of OpenAPI specification}
\end{table}

As a final signal, we bias the recommendation to retrieve higher quality specification components over lower quality ones. Every OpenAPI specification in the database was assigned a numeric quality score based on the OpenAPI 2.0 best practices~\cite{swagger}. Specifically, both the \emph{`info'} and \emph{`paths'} section of the specification were graded. For both sections, the scoring is calculated on whether it contains all the required keys, and whether the data types of the values held in each expected key matched those outlined in Table \ref{tab:quality_table}. Note that required keys are child keys that must be contained within the parent object, and expected keys are all the valid keys that could be contained within the parent object. The scores are computed using Equations~\ref{eq:quality1}-\ref{eq:quality2}.

\begin{equation}
    \label{eq:quality1}
   f_{q}(e_{i}) = \lambda_{1}q(e^{paths}_{i}) +  \lambda_{2}q(e^{info}_{i})
\end{equation}

\noindent{We empirically set $\lambda_{1}=0.7, \lambda_{2}=0.3$ in our experiments to give a higher weighting on the quality of the paths. $q(e^{paths}_{i})$ and $q(e^{info}_{i})$ are computed using Equation~\ref{eq:quality3}-\ref{eq:quality2}. Equation~\ref{eq:quality2} calculates the quality of the $info$ and $paths$ fields at depth $d$. The quality at depth $d$ is the mean of the quality at depth $d{-}1$. Equation~\ref{eq:quality3} computes the quality of $info$ and $paths$ at the lowest depth of 1. Note that for $q(e^{info}_{i})$, quality was computed up to depth 1 (\ie based on keys and values contained directly in the $info$ field) and thus only Equation~ \ref{eq:quality3} was needed to compute the quality score. For $q(e^{paths}_{i})$, quality was computed up to depth 3. More specifically, Equation~\ref{eq:quality3} was used to compute the quality of each operation that contained responses with descriptions (depth 1). Then the qualities of the operations were averaged to compute the quality of an endpoint path (depth 2), and then the qualities of the endpoints were averaged to compute $q(e^{paths}_{i})$ (depth 3) using Equation~\ref{eq:quality2}.}

\begin{equation}
    \label{eq:quality3}
   q(e^{1}_{i}) = \left\{\begin{array}{ll}
    0 & \text{if } \#(e^{1},K_{required}) < N_{K_{required}}  \\
    \frac{\#(e^{1},K_{expected},T)}{\#(e^{1}, K_{expected})} & \text{if } \#(e^{1},K_{required}) = N_{K_{required}}  \\
    \end{array}\right.
\end{equation} 

\noindent{where $\#(e^{1},K_{expected},T)$ counts the number of times the data type of value corresponding to expected keys (\ie keys that are in $K_{expected}$) present in $e^{1}$ match the expected data types in $T$, $\#(e^{1},K_{required})$ counts the number of required keys in set $K_{required}$, $\#(e^{1}, K_{expected})$ is the total number of expected keys that are present in $e^{1}$ and $N_{K_{required}}$ counts the total number of required keys. } 

\begin{equation}
    \label{eq:quality2}
    q(e^{d}_{i}) = \overline{q(e^{d-1}_{i})}, \hspace{0.5cm} \text{if  d $>$ 1}
\end{equation}

\noindent{where $d$ is the depth of the current component with respect to the original field (\ie $info$ and $paths$), and $\overline{q(e^{d-1}_{i})}$ is the mean of all $q(e_{i})$ at depth $d{-}1$.}

\section{Evaluation}\label{sec:eval}

\subsection{Dataset Availability}

The dataset was obtained from the APIS.guru Github repository \cite{apisguru}, which contains publicly available OpenAPI specifications. At the time of experimentation, this repository contained a total of 3,699 OpenAPI specifications. For the purposes of this paper, only 3,255 specifications written in OpenAPI version 2.0 were added to our dataset and given a quality score from 0 to 1. Quality scores range from 0.3 to 1, and reported a mean score of 0.85. A total of 11,778 unique endpoints were extracted from the dataset to obtain unique endpoint specifications. Each endpoint specification, along with the key components derived from its tree representation (\ie tree path tokens, text), were stored in our database. If multiple OpenAPI specifications contain the same endpoint name (\eg due to having multiple versions of the same API), the key components were concatenated and updated within the database. This database was then used to obtain the feature vectors to characterize every endpoint specification in the database and ultimately fit the fusion models evaluated. 
The dataset used in this study is available at this public repository: https://github.com/APIs-guru/openapi-directory. This public repository an open-sourced, community driven project that aims to gather a comprehensive, standards-compliant and up-to-date directory of machine-readable OpenAPI specifications.

\subsection{Models Evaluated}


Featurization methods explored for each similarity signal are outlined in Table \ref{tab:model_configs}. Note that every combination of these featurization methods was evaluated to identify the optimal configuration of the fusion model. Of all the unique fusion models that were explored, several models that resembled existing works were chosen as baseline models. To emulate works that leveraged fuzzy matching, the model which computes only the fuzzy matching of the endpoint names to evaluate similarity was chosen as a baseline. To emulate works that leveraged semantic similarity of the natural language texts present in the specification with pre-trained language models, models that only use BERT or SENT\_BERT to represent entire natural language texts were chosen as baselines. 

\begin{table}[H]
\begin{tabular}{p{3cm}|p{4.8cm}}
\hline
\hline
\textbf{Similarity approach}             & \textbf{Featurization method} \\ \hline
Tree path                                & {[} TFIDF, PPMI {]}\\ \hline
Text                                     & {[} TFIDF, PPMI, BERT, SENT\_BERT, TFIDF + BERT,                                           TFIDF + SENT\_BERT, PPMI + BERT, PPMI + SENT\_BERT {]}\\  \hline
Fuzzy                                    & {N/A}\\  

\hline
\hline

\end{tabular}
\caption{\label{tab:model_configs} Featurization methods explored}
\end{table}

\subsection{Retrieval Tasks}

\begin{figure*}[htp]
    \centering
    \includegraphics[width=17.5 cm]{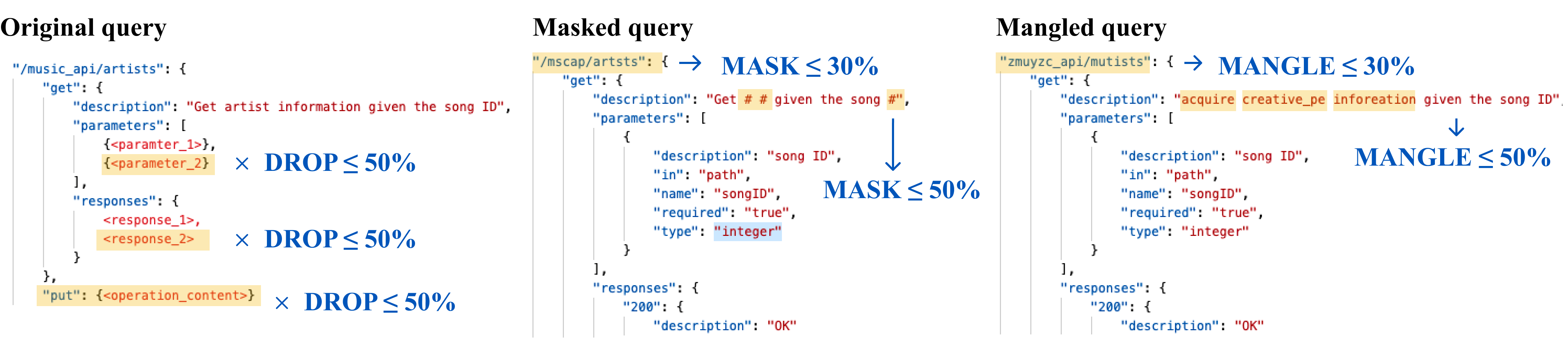}
    \caption{Modification of an endpoint specification for retrieval tasks}
    \label{fig:modification_query}
\end{figure*}

\subsubsection{Experiment setup}

For evaluation, 1000 randomly selected endpoints specifications were modified and used as queries. Recall values were then computed for each fusion model with Equation~\ref{eq:recall}:

\begin{equation}
    \label{eq:recall}
   Recall@i =  \frac{TP_i}{N}
\end{equation} 

\noindent{Where $TP\textsubscript{i}$ is the number of times the original endpoint is retrieved within its top $i = {1, 5, 10}$ recommendations, and $N$ is the total number of query samples evaluated.}

Measuring the recommendation system's ability to retrieve the original sample given a modified version of it as the query is a common evaluation method used in related works \cite{aroma}. Therefore, we evaluated API-Miner's ability to retrieve the original endpoint specification given the masked or mangled version of it. The process to obtain queries for masked and mangled retrieval is visualized in Figure \ref{fig:modification_query}, where we only visually represent the 'paths' section to ensure readability. Both masked and mangled modification begins by removing 50\% of the definitions, operations and responses present in the specification. Next, the remaining contents and the endpoint name are modified through masking or mangling. Masking involves dropping certain sections of the specification. Specifically, 50\% of the properties found under every definition were dropped and 50\% of tokens found in each operation's description or summary sections were removed. Furthermore, the endpoint names were masked by randomly removing 30\% of characters. As shown in Figure \ref{fig:modification_query}, the original endpoint name is masked and randomly selected properties and tokens are removed. For readability, the dropped tokens have been replaced with a special character (`\#'), which are eliminated during pre-processing. 

The mangling stage involves replacing certain tokens with their misspelled counterparts or its synonym (found using NLTK's WordNet library \cite{nltk}\cite{wordnet}). The misspellings were mimicked by replacing a random character of the word with another random alphabet. Specifically, 50\% of property names found under every definition were mangled and 50\% of tokens found in each operation's description or summary were mangled. Furthermore, the endpoint names were mangled by randomly replacing 30\% of characters with other random characters. As shown in figure \ref{fig:modification_query}, the original endpoint name is mangled and randomly selected tokens are replaced with misspelled counterparts (\eg information to inforeation) or its synonyms (artist to creative\_person). 

\subsubsection{Results}

\begin{table*}[t]
    \centering
    \caption{Retrieval task results}
    \begin{tabular}{*1c|*2c|*3c|*3c|*1c}
        \hline \hline
        {} & \multicolumn{2}{c}{\textbf{Model}} &  \multicolumn{3}{c}{\textbf{Masked Retrieval}} & \multicolumn{3}{c}{\textbf{Mangled Retrieval}} & \textbf{Combined}\\
        \hline
        {} & Approach & Featurization & R@1 & R@5 & R@10 & R@1 & R@5 & R@10& Average recall \\
        \hline
        \hline
        \textbf{Baseline} 
            & Text & BERT  & 0.453  & 0.555 & 0.587  & 0.379 & 0.467 & 0.497 & 0.490 \\
            & Fuzzy & N/A  & 0.426  & 0.695 & 0.782  & 0.416 & 0.689 & 0.768 & 0.629 \\
            & Text & SENT\_BERT  & 0.763  & 0.928 & 0.958  & 0.762 & 0.930 & 0.963 & 0.884 \\
        \hline
        \hline
        \multirow{3}{*}{\begin{tabular}{@{}c@{}}\textbf{Single Fusion}\end{tabular}} 
            & Tree & TFIDF & 0.505 & 0.759 & 0.830  & 0.516 & 0.768 & 0.834 & 0.702 \\
            & Text & PPMI + SENT\_BERT (t\_SVD + CCA) & 0.828 & 0.958 & 0.969 & 0.842 & 0.963 & 0.979 & 0.923 \\
        \hline
        \hline
        \multirow{3}{*}{\begin{tabular}{@{}c@{}}\textbf{Double} \\ \textbf{Fusion}\end{tabular}} 
            & \begin{tabular}{@{}c@{}}Tree, \\ Fuzzy\end{tabular} 
            & \begin{tabular}{@{}c@{}}TFIDF \\ N/A \end{tabular} & 0.692  & 0.888 & 0.932 & 0.682 & 0.877 & 0.926 & 0.833 \\
             \cline{2-10}

            & \begin{tabular}{@{}c@{}}Tree, \\ Text\end{tabular} 
            & \begin{tabular}{@{}c@{}}PPMI \\ TFIDF + BERT\_SENT (t\_SVD + CCA) \end{tabular} & 0.856 & 0.974 & 0.982 & 0.877 & 0.982 & 0.992 & 0.944 \\
             \cline{2-10}

            & \begin{tabular}{@{}c@{}}Text, \\ Fuzzy \end{tabular} 
            & \begin{tabular}{@{}c@{}}TFIDF + BERT\_SENT (t\_SVD + CCA) \\ N/A \end{tabular}  & 0.9 & 0.988 & 0.994 & 0.904 & 0.988 & 0.996 & 0.962 \\
            \cline{2-10}
        \hline
        \hline
        \multirow{3}{*}{\begin{tabular}{@{}c@{}}\textbf{Triple} \\ \textbf{Fusion}\end{tabular}} 
            & \begin{tabular}{@{}c@{}}Tree, \\ Text \\ Fuzzy \end{tabular}
            & \begin{tabular}{@{}c@{}}PPMI \\ TFIDF \\ N/A \end{tabular} & 0.896 & 0.991 & 0.995 & 0.912 & 0.993 & 0.999 & 0.964 \\
            \cline{2-10}
            
            & \begin{tabular}{@{}c@{}}Tree, \\ Text \\ Fuzzy \end{tabular}
            & \begin{tabular}{@{}c@{}}PPMI \\ TFIDF + BERT\_SENT (t\_SVD + CCA) \\ N/A \end{tabular}  & \textbf{0.910}  & \textbf{0.989} & \textbf{0.993} & \textbf{0.924} & \textbf{0.996} & \textbf{0.999} & \textbf{0.969}\\
            
        \hline
        \hline
    \end{tabular}
    \label{tab:retrieval_results}
\end{table*}
  
\begin{figure}
  \begin{subfigure}[b]{0.45\columnwidth}
    \includegraphics[width=\linewidth]{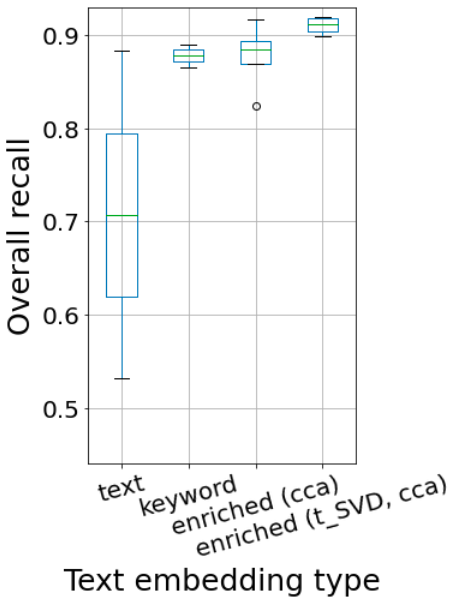}
    \caption{Masked Retrieval}
    \label{fig:1}
  \end{subfigure}
  \begin{subfigure}[b]{0.45\columnwidth}
    \includegraphics[width=\linewidth]{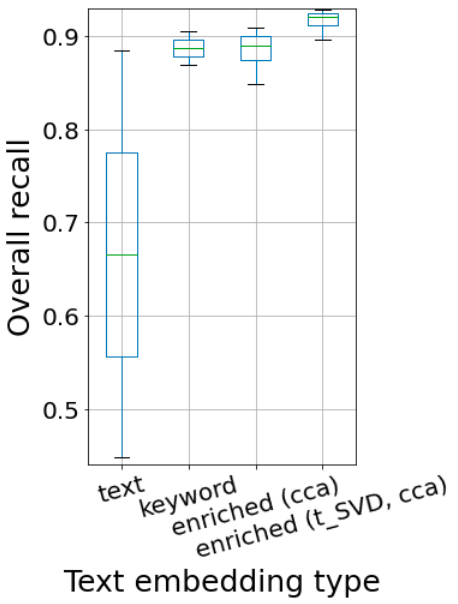}
    \caption{Mangled Retrieval}
    \label{fig:2}
  \end{subfigure}
  \caption{Overall recall obtained from models using only text features, categorized by type of text feature used. Note that there were 2 combinations of text (\ie~BERT, BERT\_SENT) and keyword embeddings (\ie~TF-IDF, PPMI) and 4 combinations of enriched embeddings evaluated.}
  \label{fig:text_features_box_plot}
\end{figure}

Table \ref{tab:retrieval_results} highlights retrieval task results obtained from top performing models by category. Upon comparing single fusion model performances, we can compare the strength of different signals (\ie tree features, text features, \etc) for retrieving relevant endpoint specifications. We observe that models using only text features perform significantly better than models using only tree or fuzzy features. We further analyze the performance of models using only text features as shown in Figure \ref{fig:text_features_box_plot} and observe that enriched representations of natural language text outperform other approaches. Specifically, we observe that applying truncated SVD to the keyword embeddings, then applying CCA to linearly project the keyword and entire text embeddings prior to concatenating yields the best results in both retrieval tasks. Overall, we observe a trend of increasing performance from single to triple fusion, with best combined recall of 0.969 achieved by a triple fusion model. This supports our hypothesis that considering multiple sources of similarity leads to better performance.

\subsection{Pooling User Study}

\subsubsection{Experiment Setup}

The pooling user study was designed to evaluate API-Miner's subjective usefulness to users who will be writing OpenAPI specifications. For this experiment, we wanted to simulate a likely scenario where the user starts developing an endpoint specification, and uses this early draft as the query to API-Miner to retrieve relevant endpoints. To obtain realistic early draft queries, we recruited 4 developers who had Master's degrees (at minimum) in quantitative disciplines and had an average of 7.5 years of industrial programming experience. By recruiting annotators with similar technical backgrounds to real users, we aimed to evaluate the tool's usefulness to potential users. Prior to the user study, the developers were provided with detailed examples to familiarise themselves with the specific version of the OpenAPI specification framework.

Each annotator was randomly assigned a task to design an endpoint by writing an early draft of an OpenAPI specification. These task prompts were automatically generated by selecting random endpoints from the database and extracting: (1) API summary (\ie~``title'', ``summary'' and ``description'' from the ``info'' section of the specification), (2) the task description (\ie~"description" and "summary" of the endpoint's operations). To allow the annotators to have enough information to properly design their queries, only tasks that contained at-least 10 tokens in the API summary and task description were considered. Annotators were instructed to design at least one of the tasks(\ie~operation descriptions) in the prompt. 

The top 10 performing models that achieved the best average recall from retrieval tasks, along with baseline models, were evaluated for the user study. Given an early draft query written by the annotator, each model generated its top 5 recommendations, resulting in a pool of recommended endpoints per query. The recommendation system's usefulness is defined by its ability to retrieve useful endpoint specifications given the user query. We defined usefulness into 4 categories: (0) non-useful, (1) minimally useful, (2) slightly useful and (3) highly useful. An endpoint rated 0 is one that has nothing to do with the functionality or content of the query. An endpoint rated 1 or 2 is one that contains functionality or topics related to the query. An endpoint rated 3 is one that contains useful information or closely matching content to the query. By using a four-point likert scale, we ensure that the annotator cannot record a neutral answer therefore distinctively classifying an endpoint to contain useful information or not. To account for annotator bias (\ie~some annotators only give high scores while others only give low scores), each of the annotator's responses were normalized from 0 to 1. Then a threshold of 0.5 was set to convert annotator scores into binary labels of 0 (not useful to the query) and 1 (useful to the query). In total, there were 846 samples (\ie~query - retrieved endpoint pairs) annotated in a pooling strategy across annotators, allowing us to calculate precision, recall and F1 for each model with Equations \ref{eq:precision_us} - \ref{eq:f1_us}:

\begin{equation}
    \label{eq:precision_us}
   Precision = \frac{TP}{TP + FP}
\end{equation} 

\begin{equation}
    \label{eq:recall_us}
   Recall = \frac{TP}{TP+FN}
\end{equation} 

\begin{equation}
    \label{eq:f1_us}
   F1 = \frac{2 \times Precision \times Recall}{Precision + Recall}
\end{equation} 

\noindent{Where $TP$ is number of useful endpoints recommended by the model, $TP + FP$ is total number of endpoints recommended by the model, and $TP + FN$ is total number of useful endpoints in the pool.}

To ensure sufficient annotator agreement, we synthetically generated 5 early draft queries by heavily modifying random endpoint specifications from the database. Specifically, all sections excluding the descriptions/summaries of operations and parameters were removed and 50\% of the operations and parameters were dropped. Anything greater than at-least 5 tokens resulted in 40\% of tokens in the description and summary sections of both operations and parameters being masked. Finally, only the last 30\% of the characters in the endpoint names were kept and the rest were discarded. This generated a total of 150 samples (\ie~query-retrieved endpoint pair), which we asked each annotator to annotate. Overall, the annotators had a Fleiss kappa score of 0.485, indicating that there was moderate agreement among the annotators \cite{fleisskappa}. One point to note for there only being moderate agreement is that each annotator will have a different view of what an API endpoint should look like thus a moderate agreement can be expected in this case.

\subsubsection{Results}

\begin{table*}[t]
\centering
\begin{tabular}{*1c|*2c|*3c}
\hline
\hline
{} & \textbf{Approach } & \textbf{Featurization} &  \textbf{Recall} &  \textbf{Precision} &  \textbf{F1}\\ \hline

\multirow{3}{*}{\textbf{Baseline}} 
    &  Fuzzy
    & N/A  & 0.240 & 0.154 & 0.188  \\
    \cline{2-6}
    
    &  Text
    & BERT  & 0.335  & 0.220 & 0.266  \\
    
    &  Text
    & BERT\_SENT  & \textbf{0.665} & \textbf{0.440} & \textbf{0.530} \\
    
\hline
    
\multirow{10}{*}{\textbf{API-Miner}} 
    & \begin{tabular}{@{}c@{}} Text \\ Fuzzy \end{tabular}
    & \begin{tabular}{@{}c@{}} TFIDF + BERT\_SENT (t\_SVD + CCA) \\ N/A \end{tabular}  & 0.665  & 0.451 & 0.538 \\
    \cline{2-6}
    
    & \begin{tabular}{@{}c@{}}Tree, \\ Text \\ Fuzzy \end{tabular}
    & \begin{tabular}{@{}c@{}}PPMI \\ TFIDF + BERT\_SENT (t\_SVD + CCA) \\ N/A \end{tabular}  & 0.680  & 0.453 & 0.544 \\
    \cline{2-6}
    
    & \begin{tabular}{@{}c@{}}Tree, \\ Text \\ Fuzzy \end{tabular}
    & \begin{tabular}{@{}c@{}}PPMI \\ TFIDF \\ N/A \end{tabular}  & \textbf{0.690}  & \textbf{0.475} & \textbf{0.562} \\
    \cline{2-6}
  
\hline
\hline

\end{tabular}
\caption{\label{tab:user_study_results} User study results on top 3 API-Miner models and baseline models}
\end{table*}

Table \ref{tab:user_study_results} reports the top 3 API-Miner and baseline models from the user study. We observe that user study results are consistent with the retrieval task results, in which the top 3 models from both evaluations match. Overall, the best performing API-Miner model surpasses the best performing baseline model by 2.5\% recall, 3.5\% precision and 3.2\% F1. 

\subsection{Discussion}
In retrieval tasks, we evaluated API-Miner's ability to retrieve the original endpoint given a synthetically modified version of it and discovered several interesting findings. We show that API-Miner is able to handle syntactically invalid specs, emphasising the usefulness for developers querying on the fly using API-Miner. First, we observe that natural language text contains the strongest signal for endpoint relevance. Secondly, we observe that enriched text features consistently outperform other text features across all evaluation tasks. The combination of keyword and entire text features is shown to provide an enriched representation of the natural language text present in API specification documents that accounts for low-resource / domain specific terms. Furthermore, we observe that applying truncated SVD to the sparse keyword features prior to projecting them using CCA yields the most meaningful representations of the natural language text. We observed that the performance of triple fusion models leveraging TF{-}IDF text features and TF{-}IDF + BERT\_SENT features perform similarly across all evaluation tasks. This suggests that for the scope of OpenAPI specification matching and recommendation, we benefit less from the rich representations obtained from large pre-trained language models. Overall, we conclude that the fusion of tree, text and fuzzy similarity approaches produce the best performing model for OpenAPI recommendation, supporting our proposal that a probabilistic log-linear model effectively optimizes recommendation performance. 

Throughout the user study, we evaluated API-Miner's usefulness in a realistic scenario. Given a task prompt, developers were instructed to write an early draft of an end point specification and then asked to evaluate the usefulness of the endpoints retrieved by each model. We observed that user study results are consistent with the retrieval tasks, and obtained best performance of 0.690 recall, 0.475 precision and 0.562 F1.

\section{Conclusion}\label{sec:conclusions}

\subsection{Threats to validity and limitations}
Threats to validity of our findings relate to the setup for the user study. To keep the annotation load manageable, we conducted a pooling-based user study. Although pooling is an acceptable approach for building a test collection in the field of Information Retrieval~\cite{jones1975report,Zobel98}, the evaluation may become biased (especially with bigger collection sizes)~\cite{Buckley06}. Furthermore, it is difficult to obtain common consensus in API annotation due to challenges such as varying developer preferences. This resulted in relatively low/moderate agreement among the annotators in our user study (\ie~kappa = 0.485). To mitigate the limitations of the user study, we also considered an automated method of evaluation (\ie~retrieval tasks).

\subsection{Summary}

This paper outlines the core principles of API-Miner and demonstrates the system's ability to recommend relevant and high quality OpenAPI specification components for users. We report strong performance in both retrieval tasks and user studies, showing that the optimal configuration of API-Miner can outperform the baseline by 15.4\% Recall@1 and 3.2\% F1 score respectively. By introducing well-defined methods of parsing and extracting features from OpenAPI specifications and leveraging a novel linear-late fusion similarity algorithm to retrieve relevant and high quality OpenAPI specifications, API-Miner presents pioneering research in the field of OpenAPI specification recommendation. 
We are eager to build upon our current research and expand it in several ways. Firstly, we would like to develop machine learning algorithms to learn the optimal weights for the fusion model. Rather than using grid-search to set the weights of the similarity approaches present in the model, we could learn the optimal weights using hill climbing optimisation techniques and further improve the recommendation performance. Furthermore, one of the limitations of the current implementation of API-Miner is that it only accommodates OpenAPI version 2.0. We would like to develop a more version-agnostic pre-processing module that is able to process different versions of OpenAPI specifications, and potentially even different types of web service documents. 

\bibliographystyle{IEEEtran}
\bibliography{IEEEabrv,references}

\end{document}